\documentclass{ws-ijmpa}
\usepackage[super,compress]{cite}
\usepackage{graphicx}
\usepackage{float}  
\usepackage{bm}
\usepackage{slashed}
\begin{document}
\markboth{Wei Lu}{Symmetry breaking and cosmological constant}

%
\catchline{}{}{}{}{}
%

\title{Dynamical Symmetry Breaking and Negative Cosmological Constant}

\author{Wei Lu}

\address{Manhasset, NY 11030, USA\\
weiluphys@yahoo.com}

\maketitle

\begin{history}
\received{Day Month Year}
\revised{Day Month Year}
\end{history}

\begin{abstract}
In the context of Clifford functional integral formalism, we revisit the Nambu-Jona-Lasinio-type dynamical symmetry breaking model and examine the properties of the dynamically generated composite bosons. Given that the model with 4-fermion interactions is nonrenormalizable in the traditional sense, the aim is to gain insight into the divergent integrals without resorting to explicit regularization. We impose a restriction on the linearly divergent primitive integrals, thus resolving the long-standing issue of momentum routing ambiguity associated with fermion-antifermion condensations. The removal of the ambiguity paves the way for the possible calculation of the true ratio of Higgs boson mass to top quark mass in the top condensation model. In this paper, we also investigate the negative vacuum energy resulted from dynamical symmetry breaking and its cosmological implications. In the framework of modified Einstein-Cartan gravity, it is demonstrated that the late-time acceleration is driven by a novel way of embedding the Hubble parameter into the Friedmann equation via an interpolation function, whereas the dynamically generated negative cosmological constant only plays a minor role for the current epoch. Two cosmic scenarios are proposed, with one of which suggesting that the universe may have been evolving from an everlasting coasting state towards the accelerating era characterized by the deceleration parameter approaching -0.5 at low redshift. One inevitable outcome of the modified Friedmannian cosmology is that the directly measured local Hubble parameter should in general be larger than the Hubble parameter calibrated from the conventional Friedmann equation. This Hubble tension becomes more pronounced when the Hubble parameter is comparable or less than a characteristic Hubble scale. 

\end{abstract}

\ccode{PACS numbers:11.30.Qc, 12.60.Rc, 04.50.Kd, 95.36.+x}



\section{Introduction}	
\label{sec:intro}

The discovery of the Higgs boson~\cite{H125A,H125C} has renewed the interest in possible explanations for the electroweak naturalness problem: the perturbative quantum corrections tend to draw the mass of a fundamental Higgs boson towards higher scales. One way of dealing with the naturalness problem is to regard the Higgs sector as an effective description of the low energy physics represented by a composite boson. It is conjectured that the fundamental Higgs boson can be replaced by a dynamically generated composite boson as the result of fermion-antifermion condensation. The top condensation model~\cite{TOP1,TOP2,TOP3,TOP4} has been extensively studied along this line of thinking, motivated by the proximity of top quark mass scale and the electroweak symmetry breaking scale. 

One of the challenges facing the top condensation model (for a review see Ref.~\citen{TOP5}) is to account for the vast range of fermion masses which span five orders of magnitude. Dimensionless ratios between parameters appearing in a physical theory cannot be accidentally small. This naturalness principle is elegantly defined by 't Hooft~\cite{TH}: a quantity should be small only if the underlying theory becomes more symmetric as that quantity tends to zero. Weakly broken symmetry ensures that the smallness of a parameter is preserved against possible perturbative disturbances. With a view toward explaining the fermion mass hierarchies in the context of composite electroweak Higgs bosons, we proposed the extended top condensation model in our previous work~\cite{WL4}. In addition to top quark condensation, the model involves tau neutrino and tau lepton condensations as well. The approach is based on the framework of Clifford algebra $C\!\ell_{0,6}$~\cite{WL1} (note that there is an interesting connection between Clifford algebra $\mathbb{C}l(6)$ and octonions via left-action maps ~\cite{Fur}), whereby standard model fermions are represented by algebraic spinors. There are two global chiral symmetries on top of the local gauge symmetries. The chiral symmetries are dynamically broken by 4-fermion condensations. In accordance with the naturalness principle, the chiral symmetries play a pivotal role in establishing the relative magnitudes of 4-fermion condensations, and consequently giving rise to fermion mass hierarchies.

That being said, there are still subjects related to the top condensation model that have not been explored in our earlier paper. Particularly, for composite Higgs boson models with nonrenormalizable 4-fermion interactions, there is a long-standing issue of the momentum routing ambiguity associated with the fermion bubble diagram~\cite{Will}. When a Feynman integral is convergent or logarithmically divergent, the integral is independent of the momentum routing parameter, because the parameter can be shifted away by a translation of the integration variable. When it comes to integrals that are more than logarithmically divergent, one should proceed with caution because the seemingly harmless momentum shifting changes the integral values. One quintessential example is the triangle diagrams of the Adler-Bell-Jackiw (ABJ) anomaly~\cite{ABJ1,ABJ2}, where the integrals are linearly divergent. The ambiguity is fixed by enforcing the vector Ward identity, at the expense of the axial Ward identity. 

The Feynman integral corresponding to the fermion bubble diagram is quadratically divergent~\cite{NJL, Will} and one shall seek a different way of removing the momentum routing ambiguity. This paper is an attempt of addressing this issue without going through explicit regularization. Via straightforward algebraic manipulations, we can show that momentum shifting changes a quadratically divergent integral by, amongst others,
\begin{align}
\label{eq:I_lin}
I_{lin}^{\mu} &= \int \frac{d^4 l}{(2\pi)^4}\, \frac{l^{\mu}}{(l^2-m^2)^2}.
\end{align}
This primitive integral is independent of external momentum. Since the integrand of $I_{lin}^{\mu}$ is odd in $l$, one might suppose that it vanishes upon symmetrical integration. Considering that $I_{lin}^{\mu}$ is linearly divergent, we can no longer make the cavalier assumption that $I_{lin}^{\mu}$ can be discarded. Analogous to the situation of the ABJ anomaly, the surface terms could spoil the identity $I_{lin}^{\mu} = 0$. The solution to this issue is to require that the physical outcome of a model should not depend on the ill-defined non-Lorentz-invariant primitive integrals which are more than logarithmically divergent. In other words, integrals like $I_{lin}^{\mu}$ should not show up in the final calculation results. This condition pins down the value of the momentum routing parameter, thus removing the routing ambiguity.

With the ambiguity out of the way, we turn to the other aspects of the composite boson model and its cosmological implications. It is argued that the bare fermion mass and the bare cosmological constant Lagrangian terms are prohibited by invoking global symmetries. The dynamically generated effective masses and negative vacuum energy could therefore be naturally small, due to the protection from weakly broken symmetries. The vacuum energy-related quartically divergent primitive integral is deemed as a separate parameter, in addition to the logarithmically and quadratically divergent primitive integrals. As a consequence, the vacuum energy could be decoupled from the emergent mass scale. It's also shown that the late-time cosmic acceleration can be realized in the framework of modified Friedmannian cosmology~\cite{WL2}, even in the presence of a small negative cosmological constant arising from dynamical symmetry breaking. 

The present paper is in a sense a continuation of our previous research on the extended top condensation model~\cite{WL4}. Rather than analyzing the dynamical electroweak symmetry breaking in its full extent, we will investigate the basic Nambu-Jona-Lasinio-type (NJL-type) model~\cite{NJL}. The goal is to contemplate the viability and cosmological consequences of a minimal dynamical symmetry breaking model with nonrenormalizable 4-fermion interactions. We hope that the lessons learned here could shed some light on the future study of more sophisticated models such as the extended top condensation model.

This paper is structured as follows: Section~\ref{sec:model} introduces the 4-fermion interaction model and Clifford functional integral formalism. In section~\ref{sec:integrals}, we study divergent integrals, the momentum routing ambiguity, and the negative vacuum energy. In section~\ref{sec:cosmology}, we examine the cosmological implications of a negative cosmological constant based on the modified Friedmann equation, and explore the possibility of time-varying emergent quantities. In the last section we draw our conclusions. Throughout this paper, we adopt the Planck units: $c = \hbar = G = 1$.

\section{The Dynamical Symmetry Breaking Model with 4-fermion Interactions}
\label{sec:model}
\subsection{Clifford algebra and the fermion Lagrangian}
\label{subsec:algeb}
Encouraged by the successful employment of Clifford algebra $C\!\ell_{0,6}$ in the extended top condensation model, we are going to use Clifford algebra extensively in this paper. Clifford algebra, also known as geometric algebra, is a powerful mathematical tool with various applications in physics~\cite{HEST1,HEST2,PAV, PAV2,Loun,DORA,Vaz}. Instead of $C\!\ell_{0,6}$, here we will focus on the familiar $C\!\ell_{1,3}$, which is also called Dirac algebra or spacetime algebra. The Clifford algebra $C\!\ell_{1,3}$ is defined by the vector basis $\{\gamma_{\mu}; \mu= 0, 1, 2, 3\}$ satisfying
\begin{align}
\label{eq:gamma}
&\gamma_{\mu}\gamma_{\nu}+\gamma_{\nu}\gamma_{\mu} = 2\eta_{\mu\nu},
\end{align}
where $\eta_{\mu\nu} = diag(1, -1, -1, -1)$.

An algebraic spinor $\psi(x)$ is a linear combination of all $2^{4}=16$ basis elements of Clifford algebra $C\!\ell_{1,3}$
\begin{align}
\label{eq:psi}
\psi(x) =&\psi_0(x) + \psi_1^{\mu}(x)\gamma_\mu + \frac{1}{2}\psi_2^{\mu\nu}(x)\gamma_{\mu}\gamma_{\nu} \nonumber\\
			&+ \psi_3^{\mu}(x)i\gamma_{\mu} + \psi_4(x) i,
\end{align}
where  $\psi_2^{\mu\nu}(x) = - \psi_2^{\nu\mu}(x)$ and the unit pseudoscalar 
\begin{align*}
i = \gamma_0\gamma_{1}\gamma_{2}\gamma_{3}
\end{align*}
squares to $-1$, anticommutes with Clifford-odd elements, and commutes with Clifford-even elements. Due to the fermion nature, the $16$ linear combination coefficients such as $\psi_0(x)$, $\psi_1^{\mu}(x)$, etc. are real Grassmann numbers. It is worth noting that the algebraic spinor as expressed in eq.~\ref{eq:psi} should not be confused with a bispinor, which is effectively bosonic and can be also expanded in terms of the 16 elements of $C\!\ell_{1,3}$. The interested readers shall refer to Refs.~\citen{HEST1,HEST2,PAV,Loun,DORA,Vaz} and especially section 4.1 in Ref.~\citen{PAV2} for detailed expositions on the mapping between an algebraic spinor and a conventional column spinor. 

Spinors with left (right) chirality correspond to Clifford-odd (even) multivectors
\begin{align*}
\psi &= \psi_{L} + \psi_{R}, \\
\psi_{L} &= \frac{1}{2}(\psi+ i\psi i), \\
\psi_{R} &= \frac{1}{2}(\psi - i\psi i).
\end{align*}

According to the conventional column representation of fermions, each Dirac fermion has $4$ components which correspond to $4$ complex or $8$ real Grassmann numbers. So there is an issue of reconciling the column spinor with the algebraic spinor endowed with $N = 16$ degrees of real Grassmann freedom. This issue is resolved by the seminal paper by Hestenes~\cite{HEST3}. It is pointed out that the algebraic spinor of $C\!\ell_{1,3}$ can be identified with the neutrino-electron isospin doublet. The spinor $\psi $ represents a pair of orthogonal left ideals
\begin{align*}
&\psi = \psi_{\nu} + \psi_{e},
\end{align*}
where $\psi_{\nu}$ ($\psi_{e}$) corresponds to the isospin up (down) projection of $\psi$
\begin{align*}
\psi_{\nu} &= \psi\frac{1 + {\gamma_3}{\gamma_0}}{2}, \\
\psi_{e} &= \psi\frac{1 - {\gamma_3}{\gamma_0}}{2}.
\end{align*}
The algebraic spinor of $C\!\ell_{1,3}$ is capable of accommodating Lagrangians which are both Lorentz invariant and electroweak gauge invariant (see Ref.~\citen{HEST3} for details). 

Here we are concerned with the NJL-type model with 4-fermion interactions. The fermion Lagrangian can be written as
\begin{align}
\label{eq:Lag}
&\mathcal{L} = \hat i\left\langle \bar{\psi}\slashed{\partial} \psi\right\rangle - \frac{1}{2N}g\big (\left\langle i\bar{\psi}\psi\right\rangle^2 + \left\langle i\bar{\psi}i\psi\right\rangle^2\big ),
\end{align}
where 
$\slashed{\partial} = \gamma^{\mu} \partial_{\mu}$ (we adopt the summation convention for repeated indices in this paper), $\gamma^{\mu} = \eta^{\mu\nu}\gamma_{\nu}$, $N=16$, $g$ is the 4-fermion coupling constant , $\left\langle \ldots\right\rangle$ stands for Clifford-scalar part of the enclosed expression, and the Dirac conjugate $\bar{\psi}$ is defined as
\begin{equation*}
\bar{\psi} = \psi^{\dagger}\gamma_0. 
\end{equation*}
Hermitian conjugate $\psi^{\dagger}$ takes the form
\begin{equation}
\label{eq:Her}
\psi^{\dagger} = \gamma_0\tilde{\psi}\gamma_0,
\end{equation}
where reversion of $\psi$, denoted $\tilde{\psi}$, reverses the order in any product of Clifford vectors. 

Note that the Hermitian conjugate in eq.~\eqref{eq:Her} is defined specifically for Clifford algebra $C\!\ell_{1,3}$. For Clifford algebra $C\!\ell_{0,6}$, the Hermitian conjugate would assume a different definition~\cite{WL1, WL4}. In the context of connecting Clifford algebra with the conventional matrix formalism, the interested readers are encouraged to consult section 5.1 in Ref.~\citen{PAV2} for a general and enlightening discussion concerning the distinction between reversion (which acts on Clifford algebra valued objects) and Hermitian conjugate (which acts on the matrices representing Clifford numbers). 

The kinetic term $\hat i\left\langle \bar{\psi}\slashed{\partial} \psi\right\rangle$ involves the mathematical imaginary number $\hat i$, which is different from Clifford algebra $C\!\ell_{1,3}$ pseudoscalar $i$. The imaginary number $\hat i$ commutes with all Clifford algebra elements. The $\hat i$ in the kinetic term is consistent with the fact that a self-energy loop diagram would yield an imaginary correction to the fermion propagator, since loop integrals would pick up an extra $\hat i$ via proper contour integral on the complex plane (or equivalently Wick rotation of time axis).  Note that $\hat i$ does not show up in the multi-fermion interaction term, and the same goes for Yang-Mills and gravitational interactions.

\begin{subequations}\label{eq:UV}
The massless Lagrangian is invariant under vector $U_V(1)$ global transformation
\begin{align}
\psi_L &\Rightarrow \psi_Le^{\alpha i} = e^{-\alpha i}\psi_L, \\
\psi_R &\Rightarrow \psi_Re^{\alpha i} = e^{\alpha i}\psi_R, 
\end{align}
\end{subequations}
and axial $U_A(1)$ global transformation \begin{subequations}\label{eq:UA}
\begin{align}
\psi_L &\Rightarrow e^{\beta i}\psi_L = \psi_Le^{-\beta i}, \\
\psi_R &\Rightarrow e^{\beta i}\psi_R = \psi_Re^{\beta i}.
\end{align}
\end{subequations}

The individual Lagrangian terms $\left\langle i\bar{\psi}\psi\right\rangle^2$ and $\left\langle i\bar{\psi}i\psi\right\rangle^2$ are not invariant under the axial transformation, albeit they are invariant in aggregation. If we add a bare mass term  
\begin{equation*}
m\hat{i}\left\langle i\bar{\psi}\psi\right\rangle
\end{equation*}
to the Lagrangian, the axial symmetry would be spoiled. The crux of the dynamical symmetry breaking mechanism is to induce an axial-symmetry-breaking effective mass term via interactions. 

Note that we can write down an NJL-type Lagrangian involving electron $\psi_{e}$ or neutrino $\psi_{\nu}$ only. The outcome of the forthcoming sections will not be qualitatively different. All one has to do is to change the value of $N$ from $N = 16$ to $N = 8$, where $N$ is the degrees of real Grassmann freedom. And for that matter, we shall regard $\psi$ as a generic spinor not necessarily tied to a specific fermion. For example, $\psi$ could be top lepton, whose fermion-antifermion condensation is of concern in the extended top condensation model~\cite{WL4}.

\subsection{Clifford functional integral and Schwinger-Dyson equation}
\label{subsec:SD}

To quantize the Lagrangian~\eqref{eq:Lag} in the framework of Clifford functional integral formalism, we have to leverage Clifford functional calculus. For our purpose here, the spinor $\psi(x)$ in~\eqref{eq:psi} is re-expressed as
\begin{align}
\label{eq:psi2}
&\psi(x) = \psi^a(x)\hat{\gamma}_a,
\end{align}
where index $a$ runs from $1$ to $N$, with $N = 16$, and each $\psi^a(x)$ is real-Grassmann valued. The operators $\hat{\gamma}_a$ span all the 16 basis elements of Clifford algebra $C\!\ell_{1,3}$
\begin{align*}
\hat{\gamma}_1 = 1,  \hat{\gamma}_2 = \gamma_0,  \hat{\gamma}_3 = \gamma_1,  \cdots,  \hat{\gamma}_{16} = i.
\end{align*}

We are interested in a Clifford functional derivative ${\delta}/{\delta \psi(x)}$ suitable for the spinor $\psi(x)$ defined in ~\eqref{eq:psi2},
\begin{align}
\label{eq:deriv}
\frac{\delta}{\delta \psi(x)} \equiv \hat{\gamma}^a \frac{\delta}{\delta\psi^a(x)},
\end{align}
where $\hat{\gamma}^a$ is the inverse of $\hat{\gamma}_a$ so that $\hat{\gamma}^a\hat{\gamma}_a = 1$ for each $a$, and ${\delta}/{\delta\psi^a(x)}$ follows the standard definition of Grassmann functional derivative. Note that we stick with the convention of always applying the Clifford functional derivative to the left of a functional. In the same vein, the Clifford functional derivative ${\delta}/{\delta \bar{\psi}(x)}$ against $\bar{\psi}(x)$ can also be defined. 

There are a few useful functional derivative properties:
\begin{align*}
\frac{\delta}{\delta \psi(x)} \psi(y) &= N \delta(x-y), \\
\frac{\delta}{\delta \psi(x)} i\psi(y) &= 0, \\
\frac{\delta}{\delta \psi(x)} \left\langle \psi(y) F\right\rangle &= \frac{\delta}{\delta \psi(x)} \left\langle \dot{\psi}(y) F\right\rangle + \frac{\delta}{\delta \psi(x)} \left\langle \psi(y) \dot{F}\right\rangle \\ &= \delta(x-y) F + \frac{\delta}{\delta \psi(x)} \left\langle \psi(y) \dot{F}\right\rangle,
\end{align*}
where $F$ is any Clifford functional and the dot on $\dot{\psi}(y)$ or $\dot{F}$ denotes functional derivative performed on the designated element only. The second property stems from the fact that the Clifford-odd portion and the Clifford-even portion of the derivative cancel out. 

Now we are ready for the quantization of the Lagrangian~\eqref{eq:Lag}. The generating functional $Z[\eta]$ can be represented as the Clifford functional integral
\begin{align}
\label{eq:Z}
Z[\eta] &= \int \mathcal{D}\psi e^{\frac{\hat{i}}{2}\int d^4x\mathcal{L}(x) + \frac{1}{2}\int d^4xd^4y\left\langle \bar{\eta}(x)\psi(x)\right\rangle \left\langle \eta(y)\bar{\psi}(y)\right\rangle}.
\end{align}
It is understood that $Z[\eta]$ is required to be normalized to $Z[0] = 1$. The Grassmann-odd sources $\eta(x)$ and $\bar{\eta}(x)$ are valued in the same Clifford space as $\psi(x)$ and $\bar{\psi}(x)$. Hence their respective Clifford functional derivatives ${\delta}/{\delta \eta(x)}$ and ${\delta}/{\delta \bar{\eta}(x)}$ can be defined in the same fashion. 

The combination of $\bar{\eta}(x)$ and $\eta(y)$ in the source term $\left\langle \bar{\eta}(x)\psi(x)\right\rangle \left\langle \eta(y)\bar{\psi}(y)\right\rangle$ should be deemed as a bilocal union. Therefore, $\eta$ and $\bar{\eta}$, and for that matter functional derivatives ${\delta}/{\delta \eta}$ and ${\delta}/{\delta \bar{\eta}}$, should always appear in pairs. It's worth mentioning that bilocal sources have been employed by two-particle irreducible (2PI) effective actions and approximation schemes~\cite{CJT} to go beyond the standard perturbative quantum field theory. 

Rather than treating $\psi(x)$ and $\bar{\psi}(x)$ as independent variables as in many textbooks, we regard them as dependent variables. The same logic applies to $\eta(x)$ and $\bar{\eta}(x)$. The extra ${1}/{2}$ factor in front of the action functional in eq.~\eqref{eq:Z} is necessary to keep the final physical results the same as those of the conventional formalism. 

\begin{subequations}\label{eq:prop}
We do not aspire to establish a mathematically rigorous definition of the above functional integration~\eqref{eq:Z}. Actually, the only property we need in this paper is that the functional integral of a total functional derivative is zero:
\begin{align}
\int \mathcal{D}\psi \frac{\delta}{\delta \psi(x)} F &= 0, \\
\int \mathcal{D}\psi \frac{\delta}{\delta \bar{\psi}(x)} F &= 0,
\end{align}
\end{subequations}
where $F$ is any functional. 

A specific application of the property ~\eqref{eq:prop},
\begin{align*}
&\int \mathcal{D}\psi \frac{\delta}{\delta \bar{\psi}(x)} [e^{\frac{\hat{i}}{2}\int d^4z\mathcal{L}(z) + \frac{1}{2}\int d^4zd^4z'\left\langle \bar{\eta}(z)\psi(z)\right\rangle \left\langle \eta(z')\bar{\psi}(z')\right\rangle}\bar{\psi}(y)] = 0,
\end{align*}
yields the Schwinger-Dyson (SD) equation 
\begin{align}
\label{eq:SD}
N\delta(x-y)Z[\eta] =& \slashed{\partial}_{x} \frac{\delta}{\delta \bar{\eta}(x)} \frac{\delta}{\delta \eta(y)}Z[\eta] \nonumber\\
&- \hat{i}\frac{g}{N}\Big [\frac{\delta}{\delta \bar{\eta}(x)} i\frac{\delta}{\delta \eta(y)} \left\langle \frac{\delta}{\delta \bar{\eta}(x)} i\frac{\delta}{\delta \eta(x)}Z[\eta]\right\rangle \nonumber\\
&+ i\frac{\delta}{\delta \bar{\eta}(x)} i\frac{\delta}{\delta \eta(y)} \left\langle i\frac{\delta}{\delta \bar{\eta}(x)} i\frac{\delta}{\delta \eta(x)}Z[\eta]\right\rangle\Big ] \nonumber\\
&+ \varepsilon \int d^4z \eta(x) \frac{\delta}{\delta \eta(y)} \left\langle \bar{\eta}(z) \frac{\delta}{\delta \bar{\eta}(z)}Z[\eta]\right\rangle, 
\end{align}
where $\varepsilon = 1$ in the source term is a dummy parameter. It's for book keeping purpose when we seek an approximate solution in the next subsection.

As mentioned earlier, functional derivatives ${\delta}/{\delta \eta}$ and ${\delta}/{\delta \bar{\eta}}$ should always show up in couples. In the above SD equation, it's understood that the closest ones are paired up.

\subsection{Bilocal source approximation}
\label{subsec:bilocal}

The SD equation~\eqref{eq:SD} is an exact functional-differential equation. In the presence of interactions, solving the SD equation is notoriously hard. The path well trodden is to find a perturbative solution, under the assumption that a certain coupling constant is small. Here we follow a non-perturbative iterative scheme dubbed as bilocal source approximation~\cite{Roch,Roch2}, which effectively treats $\varepsilon$ in the bilocal source term (the last term in the SD equation~\eqref{eq:SD}) as a series expansion parameter so that
\begin{align}
Z = Z_0 + Z_1 \varepsilon + Z_2 \varepsilon^2 + \cdots.
\end{align}
The equation of the zeroth-order approximation is
\begin{align}
\label{eq:SD0}
N\delta(x-y)Z_0 =& \slashed{\partial}_{x} \frac{\delta}{\delta \bar{\eta}(x)} \frac{\delta}{\delta \eta(y)}Z_0 \nonumber\\
&- \hat{i}\frac{g}{N}\Big [\frac{\delta}{\delta \bar{\eta}(x)} i\frac{\delta}{\delta \eta(y)} \left\langle \frac{\delta}{\delta \bar{\eta}(x)} i\frac{\delta}{\delta \eta(x)}Z_0\right\rangle \nonumber\\
&+ i\frac{\delta}{\delta \bar{\eta}(x)} i\frac{\delta}{\delta \eta(y)} \left\langle i\frac{\delta}{\delta \bar{\eta}(x)} i\frac{\delta}{\delta \eta(x)}Z_0\right\rangle\Big ]. 
\end{align}
It's equivalent to the self-consistent Hartree mean-field approximation. 

The solution to the zeroth-order equation is readily obtained as (normalized to $Z_0[0] = 1$)
\begin{align}
\label{eq:Z0}
Z_0[\eta] = e^{-\frac{1}{2}\int \frac{d^4p}{(2\pi)^4}\left\langle i\bar{\eta}(p)S(p)\eta(p)\right\rangle},
\end{align}
where $\eta(p) = \int d^4x \eta(x)e^{ip\cdot x}$ and $p\cdot x = p_\mu x^\mu$. The fermion Feynman propagator $S(p)$ is given by\footnote{The Lorentz-invariant Feynman propagator depends on the proper contour integral on the complex plane prescribed by $\hat i \epsilon$. For the rest of the paper, we do not explicitly write down $\hat i \epsilon$ in propagators for the sake of brevity.}
\begin{align}
\label{eq:S}
S(p) = \frac{1}{\slashed{p} - m + \hat{i}\epsilon},
\end{align}
where $\slashed{p} = p_\mu\gamma^\mu$, and the dynamically generated mass $m$ satisfies the gap equation 
\begin{align}
\label{eq:gap}
m &= \hat{i}g\int \frac{d^4p}{(2\pi)^4}\left\langle S(p) \right\rangle =\hat{i}g\int \frac{d^4p}{(2\pi)^4}\frac{m}{p^2 - m^2}.
\end{align}

In this paper, we concentrate on the case where the nonzero mass solution is energetically favored over the zero mass one (more on the dynamically generated negative vacuum energy in the next section). The existence of fermion-antifermion condensation is reflected in the nonzero value of $\int \frac{d^4p}{(2\pi)^4}\left\langle S(p) \right\rangle$. The emergent mass dynamically breaks the axial global symmetry~\eqref{eq:UA}. It's tantamount to adding an effective mass term $m\hat{i}\left\langle i\bar{\psi}\psi\right\rangle$ to the fermion Lagrangian~\eqref{eq:Lag}.  

Note that the zeroth-order functional equation allows for a ``complex'' mass (scalar plus pseudoscalar) 
\begin{align*}
m = |m|e^{\vartheta i} = |m|\cos(\vartheta) + i|m|\sin(\vartheta),
\end{align*}
which appears in the fermion propagator~\eqref{eq:S}. 
Nevertheless, it can be rotated into to a real mass via redefining the spinor (a $U_A(1)$ transformation~~\eqref{eq:UA})
\begin{align*}
\psi &\Rightarrow e^{-\frac{\vartheta}{2} i}\psi.
\end{align*}
So without loss of generality, we will focus on real mass only.

With the goal of studying the bosonic bound state properties of fermion-antifermion condensation, we need to go beyond the zeroth-order approximation and turn to the functional equation of the first-order approximation
\begin{align}
\label{eq:SZ1}
N\delta(x-y)Z_1 = & \slashed{\partial}_{x} \frac{\delta}{\delta \bar{\eta}(x)} \frac{\delta}{\delta \eta(y)}Z_1 \nonumber\\
&- \hat{i}\frac{g}{N}\Big [\frac{\delta}{\delta \bar{\eta}(x)} i\frac{\delta}{\delta \eta(y)} \left\langle \frac{\delta}{\delta \bar{\eta}(x)} i\frac{\delta}{\delta \eta(x)}Z_1\right\rangle \nonumber\\
&+ i\frac{\delta}{\delta \bar{\eta}(x)} i\frac{\delta}{\delta \eta(y)} \left\langle i\frac{\delta}{\delta \bar{\eta}(x)} i\frac{\delta}{\delta \eta(x)}Z_1\right\rangle\Big ]\nonumber\\
&+ \int d^4z \eta(x) \frac{\delta}{\delta \eta(y)} \left\langle \bar{\eta}(z) \frac{\delta}{\delta \bar{\eta}(z)}Z_0\right\rangle. 
\end{align}

\begin{subequations}\label{eq:Z1}
According to the above equation, the first-order generating functional can be calculated as\footnote{As an only exception to the rule of pairing up closest $\eta$ and $\bar{\eta}$, $\bar{\eta}(l)$ and $\eta(l')$ ($\bar{\eta}(l')$ and $\eta(l)$) are paired up in eq.~\eqref{eq:Z1:2}.}
\begin{align}
Z_1[\eta] =& Z_0[\eta]\Big\{ \nonumber\\
&\frac{1}{4}\int \frac{d^4l}{(2\pi)^4}\left\langle i\bar{\eta}(l)S(l)\eta(l)\right\rangle 
\int \frac{d^4l'}{(2\pi)^4}\left\langle i\bar{\eta}(l')S(l')\eta(l')\right\rangle   \label{eq:Z1:2}\nonumber\\
&+\frac{\hat{i}}{4} \int \frac{d^4p}{(2\pi)^4}D_s(p)
\Big[\int \frac{d^4l}{(2\pi)^4}\left\langle S_s\right\rangle\Big]^2 \nonumber\\
&+\frac{\hat{i}}{4} \int \frac{d^4p}{(2\pi)^4}D_p(p)
\Big[\int \frac{d^4l}{(2\pi)^4}\left\langle S_p\right\rangle\Big]^2 \nonumber\\
&+ \cdots
\Big\},
\end{align}
where
\begin{align}
S_s &= i\bar{\eta}(l-\frac{p}{2})S(l-\frac{p}{2})S(l+\frac{p}{2})\eta(l+\frac{p}{2}) \\
S_p &= i\bar{\eta}(l-\frac{p}{2})S(l-\frac{p}{2})iS(l+\frac{p}{2})\eta(l+\frac{p}{2}) 
\end{align}
\end{subequations}
The $\cdots$ terms are related to the first-order corrections to the fermion propagator $S(p)$, which we will not elaborate in this paper. 

\begin{subequations}\label{eq:D}
The effective composite boson propagators\footnote{They are generally known as fermion-antifermion channel T-matrices. We call them effective composite boson propagators conditional on the existence of fermion-antifermion condensation.} $D_s(p)$ and $D_p(p)$ in the scalar and pseudoscalar channels are 
\begin{align}
 D_s(p) &= \frac{1}{N}\frac{1}{g^{-1} - \Pi_s(p)}, \\
 D_p(p) &= \frac{1}{N}\frac{1}{g^{-1} - \Pi_p(p)}, 
\end{align}
\end{subequations}
where the bubble functions $\Pi_s(p)$ and $\Pi_p(p)$ in the scalar and pseudoscalar channels are \begin{subequations}\label{eq:Pi}
\begin{align}
\Pi_s(p) &= \hat{i}\int \frac{d^4l}{(2\pi)^4}\left\langle S(l+p)S(l)\right\rangle, \\
\Pi_p(p) &= \hat{i}\int \frac{d^4l}{(2\pi)^4}\left\langle iS(l+p)iS(l)\right\rangle,
\end{align}
\end{subequations}
which correspond to the fermion bubble diagram in the context of fermion-antifermion condensation~\cite{NJL, Will}.

The scalar and pseudoscalar propagators $D_s(p)$ and $D_p(p)$ are re-summations to infinite order chains of fermion bubble diagrams. Similar leading order calculation goes by different names such as random-phase approximation, Bethe-Salpeter T-matrix equation, or 1/N expansion. The collective modes of the composite bosons can be determined by the poles of $D_s(p)$ and $D_p(p)$. 

The integrals in the gap equation~\eqref{eq:gap} and eq.~\eqref{eq:Pi} are quadratically divergent. The analysis of divergent integrals is the focus of the next section.

\section{Divergent Integrals}
\label{sec:integrals}
\subsection{Implicit regularization}
\label{subsec:implicit}
The calculations of quantum field theory are plagued by divergent integrals, which need to be regularized at intermediate steps. After renormalization, a finite and regularization-scheme independent result can be obtained for renormalizable theories. The fermion contact interactions render the NJL-type model nonrenormalizable. Unlike the renormalizable theories, an NJL-type model depends on the form of regularization chosen, hence the regularization procedure is regarded as an integral part of the definition of the model. In the literature, the NJL model has been presented with many schemes: non-covariant 3-momentum cutoff, covariant 4-momentum cutoff in Euclidean space, and proper time regularization, among others (see Ref.~\citen{Klev} for a review).

In this paper, we will adopt the technique of implicit regularization~\cite{Batt,Batt2}. The below identity
\begin{align}
\label{eq:implicit}
\frac{1}{(l+p)^2-m^2}
=\frac{1}{(l^2-m^2)}
-\frac{p^2+2p \cdot l}{(l^2-m^2)\left[(l+p)^2-m^2\right]},
\end{align}
is applied repeatedly to the integrand of a divergent integral so that the divergent parts are isolated in primitive integrals that are independent of external physical momentum (i.e. $p$). Only the convergent integrals are allowed to involve external momentum in the denominator. Because the convergent integrals are separated from the divergent ones, the finite parts can be integrated free from effects of regularization. 

On the other hand, granted that the divergent primitive integrals are independent of external momentum, they can be treated as finite quantities as a result of unspecified regularization. It will be shown later that there are three Lorentz-invariant divergent integrals $I_{log}$, $I_{quad}$, and $I_{quar}$. They are regarded as free parameters of the model that shall be fixed (or constrained, see details in later subsections) by physical quantities, such as the dynamically generated fermion mass, composite boson mass, and vacuum energy. Explicit regularization is therefore bypassed.

As explained in the introduction section, we exercise extra caution while handling non-Lorentz-invariant divergent primitive integrals. We diverge from the standard implicit regularization approach~\cite{Batt,Batt2} when linearly or more than linearly divergent non-Lorentz-invariant integrals (such as $I^{\mu}_{lin}$ in eq.~\eqref{eq:I_lin}) are concerned. We deem these integrals as ill-defined which should not appear in the final physical results, as opposed to throwing some away based on symmetrical integration argument or stipulating the others via enforcing the consistency conditions (equivalent to discarding surface terms). 

\subsection{Gap equation}
\label{subsec:gap}
For the NJL-type model with 4-fermion interactions, the first Lorentz-invariant primitive integral we encounter is
\begin{align}
\label{eq:I_quad}
I_{quad} = \int \frac{1}{\hat{i}}\frac{d^4l}{(2\pi)^4}\frac{1}{m^2 - l^2},
\end{align}
which appears in the gap equation~\eqref{eq:gap}. This quadratically divergent integral is independent of external momentum. Note that $I_{quad}$ is real and positive, since the contour integral (with implicit regularization) on the complex plane (or equivalently Wick rotation of time axis) would pick up an extra imaginary number $\hat i$, canceling out the $\hat i$ in the denominator.

In the context of the conventional regularization scheme with a cutoff scale $M$, the gap equation implies that the coupling constant $g$ has to be fine-tuned to a value just slightly larger than a critical value $g_{crit}\sim M^{-2}$, in order to establish the hierarchy between $M$ and the much smaller fermion mass $m \ll M$.  However, given that no explicit regularization is required in our calculation, the notion of cutoff scale $M$ or critical coupling $g_{crit}(M)$ is of no relevance here. We take the view that one is agnostic of the values of the coupling constant $g$ and integral $I_{quad}$. Individually, they are left as arbitrary, insofar as they jointly meet the condition
\begin{align}
\label{eq:I_quad_g}
gI_{quad} = 1,
\end{align}
which is dictated by the gap equation~\eqref{eq:gap}. 

In accordance with 't Hooft's technical naturalness principle~\cite{TH}, once the gap equation is satisfied for a symmetry breaking scale of order $m$, it's ensured that the smallness of $m$ is preserved against possible higher order disturbances, due to the protection from the weakly broken axial symmetry. There is no need for further fine-tuning.

\subsection{Momentum routing ambiguity and composite boson mass}
\label{sec:routing}
\begin{subequations}\label{eq:Pi2}
Another quadratically divergent integral concerns the bubble functions $\Pi_s(p)$ and $\Pi_p(p)$ in eq.~\eqref{eq:Pi} corresponding to the fermion bubble diagram. To underscore the ambiguity in momentum routing, we re-express the bubble functions $\Pi_s(p)$ and $\Pi_p(p)$ in the scalar and pseudoscalar channels as
\begin{align}
\Pi_s(p) &= \hat{i}\int \frac{d^4l}{(2\pi)^4}\left\langle S(l+(1-\alpha)p)S(l-\alpha p)\right\rangle, \\
\Pi_p(p) &= \hat{i}\int \frac{d^4l}{(2\pi)^4}\left\langle iS(l+(1-\alpha)p)iS(l-\alpha p)\right\rangle,
\end{align}
\end{subequations}
where $\alpha$ is an arbitrary parameter not determined by the theory~\cite{Will}.

Unlike the case of convergent or logarithmically divergent integrals, the seemingly innocuous momentum shifting changes the integral values which are linearly or more than linearly divergent. The central question is whether the momentum routing parameter $\alpha$ is a spurious artifact arising from the approximation scheme or it's indeed a phenomenologically consequential parameter that needs to be ascertained one way or another. 

Let's begin with straightforward algebraic manipulation of $\Pi_p(p)$ in the pseudoscalar channel
\begin{align*}
&\Pi_p(p) \\
= &\hat{i}\int \frac{d^4l}{(2\pi)^4}\left\langle i\frac{1}{\slashed{l}+(1-\alpha)\slashed{p} -m}i\frac{1}{\slashed{l}-\alpha \slashed{p} -m}\right\rangle \\
= &\hat{i}\int \frac{d^4l}{(2\pi)^4}\frac{(l+(1-\alpha)p) \cdot (l-\alpha p) - m^2}{[(l+(1-\alpha)p)^2 - m^2][(l-\alpha p)^2 - m^2]} \\
= &\frac{1}{2}\hat{i}\int \frac{d^4l}{(2\pi)^4}\Big\{\frac{1}{(l-\alpha p)^2 - m^2} + \frac{1}{(l+(1-\alpha)p)^2 - m^2} \\
& -  \frac{p^2}{[(l+p)^2 - m^2][l^2 - m^2]}\Big\}.
\end{align*}
Note that we have shifted momentum routing in the last term since it's allowed for a logarithmically divergent integral. Now we apply the identity ~\eqref{eq:implicit} repeatedly to the integrands above, so that divergent integrals are isolated in terms that are independent of external momentum $p$. Collecting all the terms, we get 
\begin{align}
\label{eq:Pi_p}
\Pi_p(p) =& I_{quad} + (2\alpha -1)\hat{i}I_{lin}^{\mu} p_\mu  \nonumber\\
&+ \frac{I_{log}}{2}p^2 + \frac{1-2\alpha+2\alpha^2}{64\pi^2}p^2 +\cdots,
\end{align}
where $\cdots$ stands for finite integrals of order $O(p^4)$ (and up) and will be neglected hereafter. The quadratically divergent primitive integral $I_{quad}$ is given by ~\eqref{eq:I_quad} and it satisfies the condition ~\eqref{eq:I_quad_g}. The linearly divergent primitive integral $I_{lin}^{\mu}$ is given by~\eqref{eq:I_lin}.  The logarithmically divergent primitive integral is defined as
\begin{align}
\label{eq:I_log}
I_{log} = \int \frac{1}{\hat{i}}\frac{d^4l}{(2\pi)^4}\frac{1}{(m^2 - l^2)^2}.
\end{align}

As discussed in the introduction section, despite the fact that the integrand of $I_{lin}^{\mu}$ is odd in $l$, $I_{lin}^{\mu}$ may not necessarily vanish upon symmetrical integration. It's imperative that the physical outcome of a model should not depend on the ill-defined non-Lorentz-invariant primitive integrals like $I_{lin}^{\mu}$ which are more than logarithmically divergent. In other words, $\Pi_p(p)$ should depend on Lorentz-invariant $p^2$ only. Therefore, we are compelled to impose the condition
\begin{align}
\label{eq:condition}
(2\alpha -1)\hat{i}I_{lin}^{\mu} p_\mu &= 0.
\end{align}
This condition fixes the value of the $\alpha$ parameter to
\begin{align*}
\alpha &= \frac{1}{2}, 
\end{align*}
thus removing the momentum routing ambiguity. With that, the composite boson propagator $D_p(p)$ in the pseudoscalar channel is calculated as (utilizing the gap equation as well)
\begin{align}
\label{eq:D_p}
 D_p(p) &\approx \frac{-1}{\frac{N}{2}(I_{log} + \frac{1}{64\pi^2})p^2} = \frac{-1}{\frac{NI_{log}}{2}(1 + \Delta)p^2}\textbf{},
\end{align}
where
\begin{align*}
 \Delta &= \frac{1}{64\pi^2I_{log}}.
\end{align*}
The term $({64\pi^2})^{-1}p^2$ stems from the symmetric momentum routing (given that $\alpha = {1}/{2}$) of 
\begin{align*}
&\hat{i}\int \frac{d^4l}{(2\pi)^4}\Big\{\frac{1}{(l-p/2)^2 - m^2} + \frac{1}{(l+p/2)^2 - m^2}\Big\} \\
& = \frac{1}{64\pi^2}p^2 + 2\hat{i}\int \frac{d^4l}{(2\pi)^4}\frac{1}{l^2 - m^2}.
\end{align*}
Historically, such an extra term (proportional to $p^2$) is either treated as ambiguous~\cite{Will} or totally absent in most NJL-type calculations. Of particular note is that the pseudoscalar composite boson mass $m_p$ is still zero, unchanged by the extra term, so that the Nambu-Goldstone requirement is satisfied.

In the same fashion, the composite boson propagator $D_s(p)$ in the scalar channel can be deduced as
\begin{align}
\label{eq:D_s}
 D_s(p) &\approx \frac{-1}{\frac{NI_{log}}{2}[( 1+ \Delta)p^2 - 4m^2]}. 
\end{align}
In the presence of the extra term $\Delta p^2$ (stemming from the momentum routing fixed at $\alpha = {1}/{2}$), the pole of the scalar propagator is shifted away from $4m^2$. The ratio between the scalar composite boson mass $m_b$ and the fermion mass $m$ is given by
\begin{align}
\label{eq:ratio}
 \frac{m_b}{m} &\approx \frac{2}{\sqrt{1 + \Delta}},
\end{align}
therefore the composite boson mass $m_b$ is less than $2m$. Similar to the case of the fermion mass $m$, the dynamically generated composite boson mass (which is constrained by $m_b < 2m$ at the level of approximation) is also protected by the weakly broken axial symmetry, because both masses are contingent on dynamical symmetry breaking.

We take the view that the model's predictability depends on the emergent quantities conditional on dynamical symmetry breaking. The measurements of the fermion mass $m$ and the scalar composite boson mass $m_b$ fix the parameters of the model. Unlike the quadratically divergent counterpart $I_{quad}$, the logarithmically divergent primitive integral $I_{log}$ is dimensionless and can be inferred from the ratio $m_b/m$. The integral $I_{log}$ is thus the only determinable divergent primitive integral, whereas we only have partial knowledge of the dimensionful integral $I_{quad}$ via the constraint~\eqref{eq:I_quad_g}. 

When it comes to the top quark condensation model, one phenomenological problem is related to the conventional prediction of the Higgs-top mass ratio. Since the 2012 discovery, the Higgs boson is known to be lighter than the top quark. The top condensation model appears to fail since it gives too heavy Higgs mass compared with top quark mass. After factoring in the effects from the standard model gauge interactions and the non-leading-order corrections, the issue accompanying the top condensation model is alleviated but remains unresolved. If we take into account the momentum routing-related contributions along the lines presented in this paper, the updated calculation may possibly lead to a Higgs-top mass ratio that matches with measurements. 

\subsection{Dynamically generated vacuum energy}
\label{subsec:vacuum}
As long as quantum field theory is concerned, the absolute value of vacuum energy is normally irrelevant, because it's only the difference of energies that matters. However, gravity is coupled to the energy of the vacuum. The vacuum energy's contribution to the cosmological constant $\Lambda$ leads to measurable effects. 

For a fundamental Higgs boson, the electroweak symmetry breaking gives rise to a vacuum energy of the order
\begin{align*}
\rho_{vac} \sim \upsilon^4, 
\end{align*}
where $\upsilon$ is the electroweak scale. It is exorbitantly large ($10^{55}$ times too large) compared with the commonly accepted estimation of $\Lambda$. The cosmological constant problem is perceived as the most severe problem in physics (see Ref.~\citen{Wein} for a review). 

\begin{subequations}\label{eq:gauge}
In the context of dynamical symmetry breaking induced by multi-fermion interactions, one might wonder whether the vacuum energy's contribution to the cosmological constant is more tractable. Before answering this question, we would like to present an argument against a bare cosmological constant Lagrangian term. Let's introduce the basic building blocks of the curved spacetime in terms of Lorentz gauge theory of gravity (see Refs.~\citen{Hehl,Rand} for reviews),
\begin{align}
&e = e^{a}\gamma_{a} = e^{a}_{\mu}dx^{\mu}\gamma_{a}, \\
&\omega = \frac{1}{4}\omega^{ab}\gamma_{ab} = \frac{1}{4}\omega^{ab}_{\mu}dx^{\mu}\gamma_{ab},
\end{align}
\end{subequations}
where vierbein $e$ and spin connection $\omega$ are Clifford-valued $1$-forms, $\omega^{ab}_{\mu} = - \omega^{ba}_{\mu}$, and $\gamma_{ab} = (\gamma_{a}\gamma_{b} - \gamma_{b}\gamma_{a})/2$. The covariant derivative of the spinor field $\psi(x)$ is defined by
\begin{align}
\label{eq:Deriv}
D\psi &= (d + \omega)\psi,
\end{align}
where the spin connection $\omega$ is essential in maintaining the local Lorentz covariance of $D\psi$. The spin connection curvature $2$-form is expressed as
\begin{align}
\label{eq:R}
&R = d\omega + \omega^2 = \frac{1}{4}R^{ab}\gamma_{ab} = \frac{1}{4}(d\omega^{ab} + \eta_{cd}\omega^{ac}\omega^{db})\gamma_{ab},
\end{align}
where outer products between differential forms are implicitly assumed.

\begin{subequations}\label{eq:Lags}
The spinor kinetic, $4$-fermion interaction, gravity, and bare cosmological constant Lagrangian terms are of the forms
\begin{align}
\mathcal{L}_{spinor-kinetic} &\sim \hat{i}\left\langle \bar{\psi}ie^3 D\psi \right\rangle, \\
\mathcal{L}_{4-fermion} &\sim \left\langle ie^{4} \right\rangle(\left\langle i\bar{\psi}\psi\right\rangle^2 + \left\langle i\bar{\psi}i\psi\right\rangle^2), \label{eq:Lags:2}\\
\mathcal{L}_{gravity} &\sim \left\langle ie^{2}R \right\rangle, \\
\mathcal{L}_{bare-CC} &\sim \left\langle ie^{4} \right\rangle.
\end{align}
\end{subequations}
\begin{subequations}\label{eq:theta}
We propose a global transformation\footnote{One might define ``vierbein dimension'' as
$[e]=1$ and $[\omega]=[d]=[\psi]=[\bar\psi] = -1$. Then the bare cosmological constant Lagrangian term is ``dimensionfull'',  whereas the other Lagrangian terms are ``dimensionless'' .}
\begin{align}
e &\Rightarrow e^{\theta\hat{i}}e, \\
\omega &\Rightarrow e^{-\theta\hat{i}}\omega, \\
d &\Rightarrow e^{-\theta\hat{i}}d, \\
\psi &\Rightarrow e^{-\theta\hat{i}}\psi, \qquad \bar\psi \Rightarrow e^{-\theta\hat{i}}\bar\psi. 
\end{align}
\end{subequations}
Note that $\bar\psi$ transforms in the same way as $\psi$, since Dirac conjugate has no bearing on the imaginary number $\hat{i}$. 
It follows that the Lagrangian terms of $\mathcal{L}_{spinor-kinetic}$, $\mathcal{L}_{4-fermion}$, and $\mathcal{L}_{gravity}$ are invariant, whereas $\mathcal{L}_{bare-CC}$ transforms as
\begin{align*}
\mathcal{L}_{bare-CC} &\Rightarrow e^{4\theta\hat{i}}\mathcal{L}_{bare-CC}.
\end{align*}
If we enforce the $\theta$ transformation symmetry, then the bare cosmological constant term would be precluded. Therefore, we only allow an effective cosmological constant, which arises from the 4-fermion term ~\eqref{eq:Lags:2} conditional on dynamical symmetry breaking of the $\theta$ symmetry. 

For the flat spacetime fermion Lagrangian ~\eqref{eq:Lag}, the vacuum potential energy can be calculated as the expectation value of the 4-fermion interaction term 
\begin{align}
\label{eq:vacuum}
V =&  \int \mathcal{D}\psi \frac{1}{4N}g(\left\langle i\bar{\psi}\psi\right\rangle^2 + \left\langle i\bar{\psi}i\psi\right\rangle^2) e^{\frac{\hat{i}}{2}\int d^4x\mathcal{L}(x)} \nonumber\\
\approx &\frac{1}{4N}g\Big(\left\langle \frac{\delta}{\delta \bar{\eta}(x)} i\frac{\delta}{\delta \eta(y)} \left\langle \frac{\delta}{\delta \bar{\eta}(x)} i\frac{\delta}{\delta \eta(x)}Z_0\right\rangle \right\rangle \nonumber\\
& + \left\langle i\frac{\delta}{\delta \bar{\eta}(x)} i\frac{\delta}{\delta \eta(y)} \left\langle i\frac{\delta}{\delta \bar{\eta}(x)} i\frac{\delta}{\delta \eta(x)}Z_0\right\rangle \right\rangle\Big)\Big|_{\eta(x)=0} \nonumber\\
=& -\frac{N}{4}m^2gI_{quar},
\end{align}
where the zeroth-order generating function $Z_0(\eta)$~\eqref{eq:Z0} is utilized. The quartically divergent primitive integral
\begin{align}
\label{eq:I_quar}
I_{quar} = \int \int \frac{1}{\hat{i}}\frac{d^4l_1}{(2\pi)^4}\frac{1}{\hat{i}}\frac{d^4l_2}{(2\pi)^4}\frac{1}{(m^2 - l_1^2)(m^2 - l_2^2)},
\end{align}
is real and positive. It's the third primitive integral, in addition to the logarithmically and quadratically divergent ones $I_{log}$ and $I_{quad}$. According to the conventional wisdom, one might erroneously expect that 
\begin{align}
\label{eq:I_quar2}
I_{quar} = I^2_{quad}.
\end{align}
The underlying assumption is that the order of integration $d^4l_1$ and $d^4l_2$ does not make a difference and one can integrate individually. This sort of reasoning  is problematic for quartically divergent integrals, since the change of integral (and/or differential) order is generally not permitted for integrals that are more than logarithmically divergent.  

As a rule of thumb, relationships involving multiplication of multiple divergent integrals such as eq.~\eqref{eq:I_quar2} should be avoided. Instead, the integral $I_{quar}$ shall be treated as an independent quantity, unrelated to $I_{quad}$ or $I_{log}$. Consequently, the presumably small value of vacuum energy can be detached from the emergent fermion/boson mass scale. Note that this small value is preserved against possible perturbations, thanks to the protection from the weakly broken $\theta$ symmetry~\eqref{eq:theta}.

We know that there could be a couple of condensations in the extended top condensation model~\cite{WL4}. And there are also quark-antiquark condensations (manifested as mesons) induced by the strong interaction in QCD. Similar to the situation of the quadratically divergent counterpart, we are agnostic of the individual magnitude of each coupling constant $g_k$ (one can only determine the ratios between $g_k$) or each integral $I_{quar}(m_k)$ (associated with the dynamically generated fermion mass $m_k$). They are left as arbitrary, albeit in aggregation they are constrained by the dynamically induced cosmological constant
\begin{align}
\label{eq:Lambda}
\Lambda &= 8\pi \sum_k{V_k} = -2\pi N \sum_k{m_k^2g_kI_{quar}(m_k)}.
\end{align}

In accordance with expectation, the vacuum with dynamical symmetry breaking is energetically favored over a symmetric one, since a negative effective vacuum potential~\eqref{eq:vacuum} is generated. 

\section{Cosmology with a Negative Cosmological Constant}
\label{sec:cosmology}
\subsection{Einstein-Cartan equations and Friedmannian cosmology}
\label{subsec:EC}
As we learned from last section, dynamical symmetry breaking yields an effective cosmological constant which could be sufficiently small. However, a disturbing fact is that it's negative, hence of the ``wrong sign''. It is widely believed that a small and positive cosmological constant can account for the observation that the expansion of the universe is accelerating~\cite{Riess1,Perl}. This picture of the $\Lambda$CDM model is based on Einstein's theory of general relativity. In this subsection, we will briefly walk through the derivations of gravity equations and the Friedmannian cosmology. In later subsections, we will discuss how to reconcile a negative cosmological constant with the accelerated expansion of the universe.

As mentioned earlier, gravity can be formulated as a Lorentz gauge theory in terms of the vierbein (or tetrad/co-frame) $e$ and the spin connection $\omega$~\eqref{eq:gauge}. The gauge approach to gravity is also known as Einstein-Cartan gravity. The spin connection $\omega$, associated with the local Lorentz group SO(1, 3)\footnote{In fact, the gauge group is the double cover of Lorentz group, namely Spin(1, 3).}, plays the role of the  gauge fields in Yang-Mills theory.

The Einstein-Cartan action of gravity (with a nonzero cosmological constant $\Lambda$) is of the form
\begin{align}
\label{eq:gravity}
\mathcal{S}_{gravity} &= \frac{1}{8\pi}\int{\left\langle i(e^{2}R - \frac{\Lambda}{4!}e^4) \right\rangle},
\end{align}
where $R = d\omega + \omega^2$ is the spin connection curvature $2$-form~\eqref{eq:R}. 

\begin{subequations}\label{eq:EC}
Field equations are obtained by varying the total action (gravity plus matter) with the fields $e$ and $\omega$ independently. The resultant Einstein-Cartan equations read
\begin{align}
&\frac{1}{8\pi}(Re + eR - \frac{\Lambda}{3!}e^3) = \mathbb{T}i, \label{eq:EC:1}\\
&\frac{1}{8\pi}(Te - eT) = \frac{1}{2}\mathbb{S}i, \label{eq:EC:2}
\end{align}
\end{subequations}
where $\mathbb{T}$ is energy-momentum current $3$-form, $\mathbb{S}$ is spin current $3$-form, and $T$ is torsion $2$-form 
\begin{align}
\label{eq:torsion}
T &= de + \omega e + e \omega = T^a \gamma_{a} = (de^{a} + \eta_{bc}\omega^{ab}e^{c}) \gamma_{a}.
\end{align}

When the spin-current $\mathbb{S}$ is zero, the second Einstein-Cartan equation~\eqref{eq:EC} amounts to enforcing the zero-torsion condition
\begin{align}
\label{eq:zeroT}
T &= de + \omega e + e \omega = 0,
\end{align}
which can be used to express the spin connection $\omega$ in terms of the vierbein $e$. In this case, the remaining (first) Einstein-Cartan equation can be shown to be equivalent to the ordinary Einstein field equations for gravity with a cosmological constant. 

The reason of taking this detour to Einstein-Cartan gravity will become clear in later subsections. For now, let's turn to cosmology. On the large scale, the spatially homogeneous and isotropic universe is depicted by the Friedmann-Lema\^itre-Robertson-Walker (FLRW) metric
\begin{equation}
\label{eq:FLRW}
ds^2 = dt^2 - a(t)^2\left(\frac{dr^2}{1-k r^2} + r^2d\Omega^2\right),
\end{equation}
where $\Omega^2 = d\theta^2 + sin^2\theta d\phi^2$ and $a(t)$ is the scale factor of the universe normalized to $a(t_0)=1$ at present day $t_0$. The constant curvature $k$ takes the value $k = 0$, $k>0$, or $k<0$ for a flat, closed, or open space, respectively, 

Aided by this FLRW metric, we can derive the Friedmann equation of cosmology from Einstein field equations (or equivalently the first Einstein-Cartan equation plus the zero-torsion condition),
\begin{align}
\label{eq:FE}
H^2 &\equiv \left(\frac{\dot{a}}{a}\right)^2 = \frac{8 \pi}{3} \rho + \frac{1}{3}\Lambda - \frac{k}{a^2},
\end{align}
where $H$ is the Hubble parameter, $\rho$ is the energy density, and dot stands for cosmic time derivative.

The Friedmann equation shall be supplemented with the cosmological equation of state, which implies that the energy densities of the different constituents of the universe $\rho = \sum_w \rho_w$ scale with $a(t)$ as
\begin{align}
\label{eq:EOS}
\rho_w(a) &\sim a^{-3(1+w)},
\end{align}
where the equation of state parameter $w$ is $0$ for non-relativistic matter (including visible matter and cold dark matter) and ${1}/{3}$ for radiation. We can also view the other components on the right hand side of the Friedmann equation as energy densities with $w = -1$ and $-{1}/{3}$ for the cosmological constant $\Lambda$ and curvature $k$ terms, respectively. Note that the cosmological constant is nowadays termed dark energy~\cite{Peeb,Cop}, given that its value may evolve with cosmic time. 

As per the scaling equation~\eqref{eq:EOS}, the cosmological constant $\Lambda$ shall eventually dominate over the other decaying components of the universe. Therefore, we are persuaded of the need for a positive $\Lambda$, which is capable of driving the late-time cosmic acceleration by virtue of its negative pressure. Such a narrative of Friedmannian cosmology hinges on the accuracy of general relativity. In the next subsection, however, we will investigate a challenge to Newton's Law of gravitation and general relativity. See Ref.~\citen{Bull} for a review of a wide range of problems with the $\Lambda$CDM model and general relativity.

\subsection{Modified Newtonian dynamics}
\label{subsec:MOND}
Newton's Law of gravitation, as the non-relativistic weak-field limit of general relativity, is contradicted by the observation that the visible matter of spiral galaxies cannot possibly account for the gravitational pull responsible for the galactic rotation curves. Extra dark matter is thus postulated to make up for the mass discrepancy. 

Curiously, the deviation from Newton's Law of gravitation only occurs when the acceleration is below a universal scale (in Planck units)
\begin{align*}
a_0 &\approx 10^{-61}.
\end{align*}
This phenomenon, which is observed in a vast array of galaxies, would necessitate dramatic fine-tuning of the dark matter distribution. 

As an alternative to the dark matter hypothesis, a modification of Newtonian dynamics (MOND) is proposed~\cite{MOND} to link the Newtonian acceleration $\bm{a_N}$ from the visible matter to the true acceleration $\bm{a}$
\begin{equation}
\label{eq:MOND}
\mu\left(\frac{a}{a_0}\right)\bm{a}=\bm{a_N}, 
\end{equation}
where $\mu(x)$ is an interpolation function
\begin{equation}
\label{eq:mu}
\mu(x) \rightarrow 1 \; {\rm for} \; x \gg 1 \; \qquad {\rm and} \; \qquad \mu(x) \rightarrow x \; {\rm for} \; x \ll 1.
\end{equation}
In the weak acceleration limit ($a \ll a_0$) of MOND, an object would circulate around mass $M$ with the velocity
\begin{equation*}
v^4 =a_0M,
\end{equation*}
which is independent of the radius. This result of MOND agrees well with the observed behavior in galaxies known as the baryonic Tully-Fisher relation~\cite{TF}. 

In light of MOND theory's success on the galactic scale, one might wonder what's MOND theory's implication for the $\Lambda$CDM model. To this end, we have to go beyond the non-relativistic theory of MOND. Most of the theoretical attempts (see Ref.~\citen{FM} for a review) concentrate on modifying the Einstein field equations, or equivalently modifying the first Einstein-Cartan equation. In an earlier paper~\cite{WL2} of ours, we took the road less traveled by: changing the second Einstein-Cartan equation, which amounts to altering the zero-torsion condition in the absence of spin current.

\begin{subequations}\label{eq:EC_N}
Let's rephrase Newtonian gravity as the non-relativistic weak-field limit of the first and second Einstein-Cartan equations
\begin{align}
& \partial_i \omega_0^{i0} = 4 \pi \rho, \\
& \partial_i e_0^0 - \omega_0^{i0} = 0,
\end{align}
\end{subequations}
where the spin current $\mathbb{S}$ and cosmological constant $\Lambda$ terms are assumed to be zero, and $\rho$ is mass density. In the parlance of Newtonian gravity, the Newtonian gravitational acceleration $\bm{a_N}$ is 
\begin{align*}
& a^i_N = -\partial_i V_N \equiv -\partial_i e_0^0 = -\omega_0^{i0} ,
\end{align*}
where the Newtonian gravitational potential $V_N$ satisfies
\begin{align*}
&\nabla^2 V_N = 4 \pi \rho.
\end{align*}
Now let's add one term to the second equation of \eqref{eq:EC_N}
\begin{align}
\label{eq:EC_N2}
& \partial_i e_0^0 - (1 + \sqrt{a_0/a_N})\omega_0^{i0} = 0,
\end{align}
where
\begin{align}
\label{eq:a_N}
&  a_N = \sqrt{(\omega_0^{i0}\omega_0^{i0})}.
\end{align}
The additional term $\sqrt{a_0/a_N}\omega_0^{i0}$ is negligible if $a_N \gg a_0$. In other words, Newtonian gravity is recovered in the limit $a_N \gg a_0$. On the other hand, this additional term becomes consequential when $a_N$ is comparable or less than $a_0$. One can write down the relationship between the Newtonian gravitational acceleration $a^i_N = -\omega_0^{i0}$ and the true gravitational acceleration $a^i= -\partial_i e_0^0$ as
\begin{align}
\label{eq:MOND2}
(1 +\sqrt{a_0/a_N})\bm{a_N} &= \bm{a} .
\end{align}
Expressing $\bm{a_N}$ in terms of $\bm{a}$ via inverting the above equation, we arrive at the MOND equation ~\eqref{eq:MOND}, with the interpolation function $\mu(x)$ defined in ~\eqref{eq:mu}. Therefore, MOND is the result of modifying the second Einstein-Cartan equation in its non-relativistic weak-field limit.

\subsection{Modified Friedmannian cosmology}
\label{subsec:MFC}
\begin{subequations}\label{eq:gauge_p}
To introduce a relativistic counterpart of the modification along the lines of the last subsection, we realize that one has to break the local Lorentz gauge symmetry while retaining the local spacial rotation symmetry. Hence we resort to the following partial vierbeins and partial spin connection
\begin{align}
&e_{S} = e^{j}\gamma_{j} = e^{j}_{\mu}dx^{\mu}\gamma_{j}, \\
&e_{T} = e^{0}\gamma_{0} = e^{0}_{\mu}dx^{\mu}\gamma_{0}, \\
&\omega_{T} =\frac{1}{4}(\omega^{j0}\gamma_{j0} + \omega^{0j}\gamma_{0j})  = \frac{1}{2}\omega^{j0}_{\mu}dx^{\mu}\gamma_{j0},
\end{align} 
where $j=1,2,3$.  These fields have the preferable property of transforming like a vector under local spacial rotations. 
\end{subequations}

The relativistic version of the modified second Einstein-Cartan equation can be written as~\cite{WL2}
\begin{align}
\label{eq:MEC}
&\frac{1}{8\pi}(\tilde{T}e - e\tilde{T}) = \frac{1}{2}\mathbb{S}i.
\end{align}
The modified torsion $2$-form $\tilde{T}$ is defined by
\begin{align}
\label{eq:Mtorsion}
\tilde{T} &= T + \Delta T_{Schw} + \Delta T_{FLRW},
\end{align}
where $T$ is the original torsion $2$-form~\eqref{eq:torsion}.
\begin{subequations}\label{eq:DT}
The additional terms $\Delta T_{Schw}$ and $\Delta T_{FLRW}$ are
\begin{align}
\Delta T_{Schw} &= \sqrt{a_0/|z_{Schw}|}(\omega_T e_S + e_S \omega_T), \label{eq:T:1}\\
\Delta T_{FLRW} &= \sqrt{3h_0/|z_{FLRW}|} (\omega_T e_T + e_T \omega_T), \label{eq:T:2}
\end{align}
\end{subequations}
where
\begin{align*}
z_{Schw} &= \frac{4! e^2(\omega_T e_S + e_S \omega_T)}{2e^4}, \\
z_{FLRW} &= \frac{4! e^2(\omega_T e_T + e_T \omega_T)}{2e^4},
\end{align*}
and the magnitude of a multivector $|M|$ is defined as
\begin{equation*}
|M| = \sqrt{\left\langle M^{\dagger}M \right\rangle}.
\end{equation*}
The modified torsion breaks the Lorentz symmetry, while it is covariant under spacial rotations. In the absence of spin current $\mathbb{S}$, which is the case studied in the current paper, the modified second Einstein-Cartan equation yields
\begin{align}
\label{eq:Tzero}
&T + (\Delta T_{Schw} + \Delta T_{FLRW}) = 0,
\end{align}
which means a change to the usual zero-torsion condition $T=0$, since $\Delta T_{Schw} + \Delta T_{FLRW}$ is non-zero in general. 

The first term of the torsion modification $\Delta T_{Schw}$~\eqref{eq:T:1} is relevant for the Schwarzschild metric. The factor of $\sqrt{a_0/|z_{Schw}|}$ in eq.~\eqref{eq:DT} reduces to $\sqrt{a_0/a_N}$ in the non-relativistic weak-field limit, hence eq.~\eqref{eq:EC_N2} is recovered from eq.~\eqref{eq:Tzero}. Therefore, the modification is indeed a relativistic parent theory of MOND. 

The interesting part of the torsion modification is the second term $\Delta T_{FLRW}$~\eqref{eq:T:2}, which bears striking resemblance to the first term $\Delta T_{Schw}$~\eqref{eq:T:1}. One only has to replace $e_S$ and $a_0$ with $e_T$ and $3h_0$, respectively. Unlike $\Delta T_{Schw}$, it turns out that $\Delta T_{FLRW}$ is actually relevant for the FLRW metric and $h_0$ is a characteristic Hubble scale, which is a new parameter independent of the characteristic MOND acceleration scale $a_0$. That being said, it is assumed that $h_0$ is of the same order of $a_0$ (in Planck units)
\begin{align*}
h_0 &\approx 10^{-61}.
\end{align*}

The torsion modification terms $\Delta T_{Schw}$~\eqref{eq:T:1} and $\Delta T_{FLRW}$~\eqref{eq:T:2} can be phenomenologically viewed as ``dark torsion''. They might also be interpreted as ``dark spin current'' (if the modification terms are moved to the right-hand side of the modified Einstein-Cartan equation~~\eqref{eq:MEC}), as an alternative to dark matter. 

\begin{subequations}\label{eq:MFE}
With the help of the FLRW metric~\eqref{eq:FLRW}, one can show that the modified second Einstein-Cartan equation~\eqref{eq:MEC}, in conjunction with the original first Einstein-Cartan equation~\eqref{eq:EC:1}, leads to the modified Friedmann equation~\cite{WL2}
\begin{align}
\label{eq:MFE:1}
&H^2_F = \frac{8 \pi}{3} \rho + \frac{1}{3}\Lambda - \frac{k}{a^2}, 
\end{align}
where 
\begin{align}
\label{eq:MFE:2}
&(1 + \sqrt{\frac{h_0}{H_F}})H_F = H.
\end{align}
\end{subequations}
Here $H_F$ is the Friedmann Hubble parameter and $H \equiv \dot{a}/{a}$ is the true Hubble parameter. The Friedmann Hubble parameter $H_F$ can be determined via eq.~\eqref{eq:MFE:2} as
\begin{align}
\label{eq:H_F}
&H_F = \mu\left(\frac{H}{h_0}\right)H,
\end{align}
with the interpolation function specified in eq.~\eqref{eq:mu}. Consequently, the modified Friedmann equation can be reformulated as 
\begin{align}
\label{eq:MFE2}
&\left[\mu\left(\frac{H}{h_0}\right)H\right]^2 = \frac{8 \pi}{3} \rho + \frac{1}{3}\Lambda - \frac{k}{a^2}. 
\end{align}
The modified Friedmann equation~\eqref{eq:MFE2} parallels the MOND equation~\eqref{eq:MOND} in the sense that the former replaces Hubble parameter $H$ with $\mu\left({H}/{h_0}\right)H$ in Friedmann equation, while the latter replaces acceleration $\bm{a}$ with $\mu\left({a}/{a_0}\right)\bm{a}$ in Newtonian dynamics.

According to this modified Friedmannian cosmology (MFC), the characteristic Hubble scale $h_0$ marks the boundary between the validity domains of Friedmannian cosmology and MFC. For large Hubble parameter $H \gg h_0$ (the Friedmannian regime), one has $H_F \approx H $. Therefore, the modified Friedmann equation~\eqref{eq:MFE2} is reduced to the usual Friedmann equation~\eqref{eq:FE}. On the other hand, in the limit of small
Hubble parameter $H \ll h_0$ (the deep MFC regime), $H_F$ is given by
\begin{align*}
&H_F \approx \frac{H}{h_0}H.
\end{align*}
The modified Friedmann equation then reads
\begin{align}
\label{eq:MFE3}
&\frac{1}{h_0^2}\left(\frac{\dot{a}}{a}\right)^4 = \frac{8 \pi}{3} \rho + \frac{1}{3}\Lambda - \frac{k}{a^2},
\end{align}
which departs from the conventional Friedmannian cosmology. 

Given the relationship~\eqref{eq:MFE:2} (equivalently the interpolation relationship~\eqref{eq:H_F}), the Friedmann Hubble parameter $H_F$ is generally smaller than the Hubble parameter $H$. If one attempts to calibrate the Hubble parameter via the conventional Friedmann equation~\eqref{eq:FE}, it's actually $H_F$ that is inferred, which differs from the true Hubble parameter $H = \dot{a}/{a}$. This discrepancy could be manifested in the ``$H_0$ tension'' between CMB-predicted value of the Hubble parameter~\cite{Riess2} in concert with $\Lambda$CDM (corresponding to $H_F(t_0)$) and the local measurements from supernovae~\cite{Planck} (which prefer a higher value, corresponding to $H_0=H(t_0)$). 

Considering that MOND negates the need for dark matter in galactic systems, a natural question would be whether MFC, as a relativistic parent theory of MOND, can explain the cosmological mass discrepancies without invoking cold dark matter (CDM). Our hypothesis is that MFC could potentially reduce CDM's percentage in the total mass-energy budget of the universe. Nevertheless, we leave open the possibility that there might still be some remaining CDM constituents. The prime candidates are the $4$-fermion condensations (which are electroweak singlets) in the extended top condensation model~\cite{WL4} and the sterile (right-handed) neutrinos which are endowed with seesaw-scale (believed to be much higher than the electroweak scale) Majorana masses.

\subsection{MFC scenario one: coping with negative cosmological constant}
\label{subsec:MFC1}
Now we are ready for the reconciliation of the dynamically generated negative cosmological constant $\Lambda$ with the accelerated expansion of the universe in the framework of MFC. Let's study a flat universe ($k$ = 0) after the radiation-dominated epoch. The predominant mass-energy density components are thus matter and a negative cosmological constant $\Lambda < 0$. These components can be rewritten as
\begin{align}
\label{eq:nLambda}
\frac{8 \pi}{3} \rho + \frac{1}{3}\Lambda = \bar\rho_0(a^{-3} - \lambda).
\end{align}
where $\rho$ includes both visible and dark matter, and the negative $\Lambda$ is reparametrized by the positive parameter $\lambda = -\frac{1}{3\bar\rho_0}\Lambda > 0$. We assume that the magnitude of the dynamically generated $\Lambda$~\eqref{eq:Lambda} is considerably smaller than the mass density of the current epoch, hence
\begin{align*}
\lambda \ll 1.
\end{align*}

The modified Friedmann equation~\eqref{eq:MFE2} can be cast in the form
\begin{align}
\label{eq:TV}
\dot a^2 + V(a) = 0,
\end{align}
where for $H \gg h_0$ (the Friedmannian regime)
\begin{align}
\label{eq:V_FC}
V(a) = -\bar\rho_0(a^{-1} - \lambda a^2), 
\end{align}
and for $H \ll h_0$ (the deep MFC regime)
\begin{align}
\label{eq:V_MFC}
V(a) = -h_0\sqrt{\bar\rho_0}\sqrt{(a - \lambda a^4)}.
\end{align}
This reformulation benefits from the Newtonian interpretation of a mass $m_0=2$ (with kinetic energy $T=m_0\dot a^2/2 = \dot a^2$) moving in the potential $V(a)$ subject to the constraint of zero total energy. 

Given that the Hubble parameter is large ($H \gg h_0$) at the early stage of the mass-dominated epoch, potential~\eqref{eq:V_FC} implies that (neglecting the small $\lambda$ term)
\begin{align*}
\dot{a} \sim t^{-\frac{1}{3}}, \qquad
H \approx \frac{2}{3}t^{-1}, \qquad
q \approx \frac{1}{2},
\end{align*}
where 
\begin{align*}
q = - \frac{\ddot{a}a}{\dot{a}^2}
\end{align*}
is the deceleration parameter. The positive value of $q= 1/2$ indicates that the universe is decelerating. 

With the passage of time, the ever-decreasing Hubble parameter $H \sim t^{-1}$ eventually enters the regime $H \ll h_0$. Consequently, according to \eqref{eq:V_MFC}, we have (neglecting the small $\lambda$ term)
\begin{align*}
\dot{a} \sim t^{\frac{1}{3}}, \qquad
H \approx \frac{4}{3}t^{-1}, \qquad
q \approx -\frac{1}{4}.
\end{align*}
The universe is therefore accelerating, characterized by the negative $q \approx -{1}/{4}$. We generally believe that the Hubble parameter of the current epoch is of the value
\begin{align*}
H_0 = H(t_0) \lesssim h_0.
\end{align*}
Thus we have already entered the MFC regime manifested by the late-time cosmic acceleration. 

That being said, the small and negative cosmological constant will in the end catch up with the declining matter density and become significant in the far future. The acceleration is going to give way to deceleration at the critical scale $a_{crit}$ determined by
\begin{align*}
\frac{dV(a)}{da}|_{a = a_{crit}} = 0,
\end{align*}
which yields
\begin{align}
\label{eq:a_crit}
a_{crit} = 2^{-\frac{2}{3}}\lambda^{-\frac{1}{3}}.
\end{align}
Further down the road, the expansion of the universe will grind to a halt at the maximum scale $a_{max}$ derived from
\begin{align*}
V(a)|_{a = a_{max}} = 0,
\end{align*}
which implies
\begin{align}
\label{eq:a_max}
a_{max} = \lambda^{-\frac{1}{3}} \approx 1.6 a_{crit}.
\end{align}
After reaching this maximum scale $a_{max}$, the universe will trace back and embark on contraction with a decreasing scale factor. Note that both $a_{crit}$ and $a_{max}$ are determined using the deep MFC version of potential~\eqref{eq:V_MFC}. 

\subsection{MFC scenario two: coasting universe yielding to acceleration}
\label{subsec:MFC2}
As mentioned in earlier section, according to 't Hooft's technical naturalness principal~\cite{TH}, the small scales of the dynamically generated masses and vacuum energy could be protected by the weakly broken global symmetries. Nevertheless, technical naturalness does not answer the question of why a physical quantity is small in the first place. The stronger naturalness in the sense of Dirac~\cite{Dirac1} requires that there shall be no unexplained large (or small) numbers in nature. Dirac suggested, in his large number hypothesis, that very large (or small) dimensionless ratios should be considered as variable parameters pertaining to the state of the universe. 

Inspired by Dirac's hypothesis, we propose the notion of Planck naturalness: an emergent quantity, which varies with cosmic time, shall be of order 1 at Planck time $t_p = 1$ and its current large (or small) value is determined by the age of the universe $t_0 \approx 10^{61}$ (in Planck units).

Two points should be clarified regarding Planck naturalness. First of all, we will not speculate about the universe before Planck time $t_p$, which belongs to the realm of quantum gravity. And from an effective field theory point of view, an infinite number of terms allowed by symmetry requirements should be included in a generalized action of the world. The gravity and Yang-Mills actions are the first few order terms\cite{WL1} that are relevant in the low-energy limit. Therefore, Planck time $t_p$ is the furthest point at which we may give limited credence to the low-energy effective field theory. 

Secondly, by emergent quantities we mean dynamically generated quantities, such as the fermion and composite boson masses, along with the negative vacuum energy. In comparison, the Planck units (speed of light $c$, reduced Planck constant $\hbar$, and gravitational constant $G$) are fixed. As such, the Planck units are the standard yardsticks to measure the variability of the emergent dimensionful quantities.

Rather than dwelling upon individual emergent quantities, we will focus on the total mass-energy density
\begin{align*}
\bar\rho_{tot} = \frac{8 \pi}{3} \rho + \frac{1}{3}\Lambda - \frac{k}{a^2},
\end{align*}
which is the sum of the right hand side of the modified Friedmann equation ~\eqref{eq:MFE2}. That is to say, we regard every constituent of $\bar\rho_{tot}$ as emergent and time-varying. Note that $\bar\rho_{tot}$ is assumed to be positive, albeit the contribution from the dynamically generated time-varying $\Lambda$ (dark energy) is negative and the contribution from the curvature term could be either positive or negative depending on the sign of $k$. 

Let's suppose that the time-varying mass, radiation, and dark energy densities somehow work in concert, so that the aggregation of them (plus a presumably non-zero curvature term) collectively scales with $a$ as
\begin{align}
\label{eq:rho}
\bar\rho_{tot} = \frac{1}{t_0^{2}a^2} = \frac{1}{\bar a^2},
\end{align}
where $\bar{a} = t_0a$ is the re-scaled scale factor. The equation of state parameter $w$ for $\bar\rho_{tot}$ is $-1/3$. This sort of cosmic fluid was first proposed in Ref.~\citen{Kolb} and the term K-matter has been coined, because the $w=-1/3$ fluid behaves in a similar way as the curvature term $-{k}/{a^2}$ (with a negative $k$) in the empty Milne universe.

If we for a moment forget about MFC and assume that the conventional Friedmann equation~\eqref{eq:FE} are more or less accurate for the K-matter universe, we would arrive at
\begin{align}
\label{eq:coast}
\bar{a} = t, \qquad
\bar\rho_{tot} = t^{-2},\qquad
H = t^{-1}, \qquad
q = 0.
\end{align}
This coasting universe solution, characterized by the deceleration parameter $q = 0$, has the appealing property of satisfying Planck naturalness: the dependence of $\bar\rho_{tot}(t)$ on time does not involve additional large (or small) parameter and it is of the order $\bar\rho_{tot}(t_p) = t_p^{-2} = 1$ at Planck time. 

Recent years have witnessed a renewed interest~\cite{Melia1} in this $w = -1/3$ coasting universe originally investigated in Ref.~\citen{Kolb}, because it enjoys the equality
\begin{align}
\label{eq:Ht}
Ht = 1.
\end{align}
For the current epoch, we do have approximately $H_0t_0 \approx 1$ (see, however, Refs.~\citen{Ma,Wang1,Wang2} for the uncertainty in the estimation of cosmic age $t_0$). In the context of $\Lambda$CDM, the equality is an uncanny coincidence because it is only true at present time. For the $w = -1/3$ coasting universe, the equality is satisfied at all cosmic time.

Another favorable attribute of the $w = -1/3$ model is that the horizon problem is nonexistent, since it can be demonstrated~\cite{Melia2} that the opposite sides of the cosmos have remained causally connected to each other from the very first moments of the universe. 

Notwithstanding its merits, the eternally coasting universe is disfavored at low redshift. The fact that the universe is currently in a phase of accelerated expansion has been firmly established from the observations of supernovae~\cite{Riess1,Perl}, which indicates that the present value of the deceleration parameter $q(t_0)=q_0$ is markedly less than zero $q_0<0$, contrary to the claim of $q_0=0$ for the eternally coasting universe.  

Now let's suppose that the universe has entered the MFC regime (i.e. $H_0 \lesssim h_0$) at low redshift , whereas the $w = -1/3$ assumption is still valid. The modified Friedmann equation~\eqref{eq:MFE3} in the deep MFC regime takes the form
\begin{align}
\label{eq:MFE4}
&\frac{1}{h_0^2}\left(\frac{\dot{\bar{a}}}{\bar{a}}\right)^4 = \frac{1}{\bar a^2},
\end{align}
which leads to
\begin{align}
\label{eq:accel}
\bar{a} \sim t^2, \qquad
\bar\rho_{tot} \sim t^{-4},\qquad
H \approx 2t^{-1}, \qquad
q \approx -\frac{1}{2}.
\end{align}
Hence at low redshift, the deceleration parameter $q$ should have been evolving from the coasting $q=0$ towards the accelerating $q=-1/2$. This unique behavior of $q$ is distinguishable from either the $\Lambda$CDM model or the eternally coasting model, thus it could be verified via further data analysis. 

Moreover, the Hubble parameter has been transitioning from $H = t^{-1}$ (the Friedmannian regime) to $H \approx 2t^{-1}$ (the deep MFC regime), which suggests that $2 > H_0t_0 > 1$, rather than $H_0t_0 = 1$. The inequality of $t_0 > 1/H_0$ implies that the commonly quoted cosmic age (approximately $1/H_0$) might be an underestimation.  In this regard, we would like to draw attention to the observation~\cite{Ma,Wang1,Wang2} that 9 extremely old globular clusters are older than the widely accepted cosmic age.

As a last note, we shall mention that there is an unexplained coincidence: the characteristic Hubble scale $h_0$ of MFC and the characteristic acceleration scale $a_0$ of MOND are of the same order as $1/t_0$
\begin{align*}
h_0 \approx a_0 \approx \frac{1}{t_0} \approx 10^{-61}.
\end{align*}
The underlying reason for this coincidence is unknown. One possibility is that $h_0$ and $a_0$ are emergent and time-varying as well. But we just leave it at that, without further elaborating the implications. 

\section{Conclusions}
\label{sec:concl}

In the context of Clifford functional integral formalism, we revisit the NJL-type model with nonrenormalizable 4-fermion interactions. The model leads to a bosonic bound state as the result of the fermion-antifermion condensation. The goal of the paper is to gain insight into the divergent integrals without going through explicit regularization. It is argued that the physical outcome of the model should not depend on the ill-defined non-Lorentz-invariant primitive integrals which are more than logarithmically divergent. This condition removes the momentum routing ambiguity associated with the fermion-antifermion condensation. A resultant extra term shifts the pole of the scalar bosonic channel away from $4m^2$, where $m$ is the dynamically generated fermion mass.

We present an alternative view on nonrenormalizable models conditional on dynamical symmetry breaking: a model's predictability shall depend on the emergent quantities such as the fermion and boson masses along with the dynamically generated negative vacuum energy, whereas the absolute magnitude of the coupling constant is not measurable. We argue that the bare fermion mass and the bare cosmological constant Lagrangian terms are prohibited by enforcing the global axial and $\theta$ symmetries. The dynamically generated effective masses and vacuum energy could thus be protected by the weakly broken symmetries. The vacuum energy-related quartically divergent primitive integral is regarded as a separate parameter of the model independent of the quadratically and logarithmically divergent primitive integrals. The vacuum energy is therefore decoupled from the emergent mass scale. 

In the presence of a small negative cosmological constant arising from dynamical symmetry breaking, it's demonstrated that the late-time cosmic acceleration can be accommodated in the framework of modified Friedmannian cosmology (MFC). MFC is originated from modifying the second Einstein-Cartan equation, which amounts to altering the zero-torsion condition. A characteristic Hubble scale $h_0$ demarcates the boundary between the validity domains of the Friedmannian cosmology and MFC. One prediction of MFC is that the Hubble parameter calibrated from the conventional Friedmann equation is in general smaller than the local Hubble parameter inferred from supernovae observations. This ``Hubble tension'' is more pronounced when the Hubble parameter is comparable or less than the characteristic Hubble scale $h_0$.

We propose two cosmic evolution scenarios, with one of which based on the premise that the total mass-energy density of the the universe may behave like the K-matter characterized by the equation of state parameter $w= -1/3$. It follows that the universe could be coasting with a linearly increasing scale factor for the cosmic era when the Friedmannian cosmology is applicable. It has the appealing feature of satisfying Planck naturalness: the total mass-energy density is of order $1$ at Planck time. Our view is that the universe may have already entered the MFC regime ($H \lesssim h_0$). Consequently, the deceleration parameter $q$ should have been evolving from the coasting $q=0$ towards the accelerating $q=-1/2$ at low redshift. This scenario could potentially alleviate the cosmic age problem posed by the extremely old globular clusters.

\section*{Acknowledgments}
I am grateful to Matej Pavsic and Ziqi Yan for helpful correspondences.


\begin{thebibliography}{0}    

\bibitem{H125A} ATLAS Collaboration, (G. Aad {\it et al.}), {\it Phys. Lett. B }{\bf 716}, 1 (2012).

\bibitem{H125C} CMS Collaboration, (S. Chatrchyan {\it et al.}),  {\it Phys. Lett. B }{\bf 716}, 30 (2012).

\bibitem{TOP1} Y. Nambu, in {\it New Theories in Physics, Proceedings of the XI International Symposium on Elementary Particle Physics}, eds. Z. Ajduk, S. Pokorski, and A. Trautman (World Scientific, Singapore, 1988), p.~1.

\bibitem{TOP2} V. A. Miransky, M. Tanabashi and K. Yamawaki, {\it Phys. Lett. B }{\bf 221}, 177 (1989).

\bibitem{TOP3} W. J. Marciano, {\it Phys. Rev. Lett. }{\bf 62}, 2793 (1989). 

\bibitem{TOP4} W. A. Bardeen, C. T. Hill and M. Lindner, {\it Phys. Rev. D }{\bf 41}, 1647 (1990).

\bibitem{TOP5} G. Cvetic, {\it Rev. Mod. Phys. }{\bf 71}, 513 (1999).

\bibitem{TH} G. 't Hooft, {\it NATO Adv. Study Inst. Ser. B Phys. }{\bf 59}, 135 (1980).

\bibitem{WL4} W. Lu, {\it Int. J. Mod. Phys. A }{\bf 32}, 1750159 (2017).

\bibitem{WL1} W. Lu, {\it Adv. Appl. Clifford Algebras } {\bf 21}, 145 (2011).

\bibitem{Fur} C. Furey, {\it Phys. Lett. B }{\bf 785}, 84 (2018).

\bibitem{Will} R. Willey, {\it Phys. Rev. D }{\bf 48}, 2877 (1993).

\bibitem{ABJ1} S. L. Adler, {\it Phys. Rev. }{\bf 177}, {2426} (1969).

\bibitem{ABJ2} J. S. Bell and R. Jackiw, {\it Il Nuovo Cimento A }{\bf 60}, {47} (1969).
  
\bibitem{NJL} Y. Nambu and G. Jona-Lasinio, {\it Phys. Rev. }{\bf 122}, 345 (1961).

\bibitem{WL2} W. Lu,  {\it Modified Einstein-Cartan gravity and its implications for cosmology}, arXiv:1406.7555 [gr-qc].

\bibitem{HEST1} D. Hestenes,  {\it  Space-Time Algebra}, (Gordon and Breach, New York, 1966).

\bibitem{HEST2} D. Hestenes and G. Sobczyk,  {\it  Clifford algebra to geometric calculus: a unified language for mathematics and physics}, (Kluwer Academic Publishers, Dordrecht, 1984).

\bibitem{PAV} M. Pavsic,  {\it The Landscape of Theoretical Physics: A Global View. From Point Particles to the Brane World and Beyond, in Search of a Unifying Principle}, (Kluwer Academic Publishers, Dordrecht, 2001).

\bibitem{Loun} P. Lounesto,  {\it Clifford algebras and spinors}, (Cambridge University Press, Cambridge, 2001).

\bibitem{DORA} C. Doran and A. Lasenby,  {\it  Geometric Algebra for Physicists}, (Cambridge University Press, Cambridge, 2003). 

\bibitem{Vaz} J. Vaz Jr. and R. da Rocha Jr.,  {\it  An Introduction to Clifford Algebras and Spinors}, (Oxford University Press, Oxford, 2016). 

\bibitem{PAV2} M. Pavsic, {\it Int. J. Mod. Phys. A }{\bf 21}, 5905 (2006).

\bibitem{HEST3} D. Hestenes,  {\it Found. Phys. } {\bf 12}, 153 (1982).

\bibitem{CJT} J. M. Cornwall, R. Jackiw, and E. Tomboulis, {\it Phys. Rev. D }{\bf 10}, 2428 (1974).

\bibitem{Roch} V. E. Rochev, {\it J. Phys. A }{\bf 30}, 3671 (1997).

\bibitem{Roch2} R. G. Jafarov and V. E. Rochev, {\it Central Eur. J. Phys. }{\bf 2}, 367 (2004).

\bibitem{Klev} S. P. Klevansky, {\it Rev. Mod. Phys. }{\bf 64}, 649 (1992).

\bibitem{Batt} O. A. Battistel, A. L. Mota, and M. C. Nemes, {\it Mod. Phys. Lett. A }{\bf 13}, 1597 (1998).

\bibitem{Batt2} O. A. Battistel, G. Dallabona, and G. Krein, {\it Phys. Rev. D }{\bf 77}, 065025 (2008).

\bibitem{Wein} S. Weinberg, {\it Rev. Mod. Phys. }{\bf 61}, 1 (1989).

\bibitem{Hehl} F. W. Hehl, P. Von Der Heyde, G. D. Kerlick and J. M. Nester, {\it Rev. Mod. Phys. }{\bf 48}, 393 (1976).  

\bibitem{Rand} A. Randono,  {\it Gauge gravity: a forward-looking introduction}, arXiv:1010.5822 [gr-qc].

\bibitem{Riess1} Supernova Search Team, (A. G. Riess {\it et al.}), {\it Astron. J. }{\bf 116}, 1009 (1998).

\bibitem{Perl} Supernova Cosmology Project, (S. Chatrchyan {\it et al.}),  {\it Astrophys. J. }{\bf 517}, 565 (1999).

\bibitem{Peeb} P. J. E. Peebles and B. Ratra, {\it Rev. Mod. Phys. }{\bf 75}, 559 (2003).

\bibitem{Cop} E. J. Copeland, M. Sami, and S. Tsujikawa, {\it Int. J. Mod. Phys. D }{\bf 15}, 1753 (2006).

\bibitem{Bull} P. Bull {\it et al.}, {\it Phys. Dark Univ. }{\bf 12}, 56 (2016).

\bibitem{MOND} M.  Milgrom, {\it Astrophys. J. }{\bf 270}, 365 (1983).

\bibitem{TF} R. B. Tully and J. R. Fisher, {\it Astron. Astrophys. }{\bf 54}, 661 (1977).

\bibitem{FM} B. Famaey and S. McGaugh, {\it Living Rev. Rel. }{\bf 15}, 10 (2012).

\bibitem{Riess2} A. G. Riess {\it et al.}, {\it Astrophys. J. }{\bf 826}, 56 (2016).

\bibitem{Planck} P. A. R. Ade {\it et al.}, {\it Astron. Astrophys. }{\bf 594}, A13 (2016).

\bibitem{Dirac1} P. A. M. Dirac, {\it Nature }{\bf 139}, 323 (1937).

\bibitem{Kolb} E. W. Kolb, {\it Astrophys. J. }{\bf 344}, 543 (1989).

\bibitem{Melia1} F. Melia and A. Shevchuk, {\it Mon. Not. Roy. Astron. Soc. }{\bf 419}, 2579 (2012).

\bibitem{Ma} J. Ma {\it et al.}, {\it Astron. J. }{\bf 137}, 4884 (2009).

\bibitem{Wang1} S. Wang {\it et al.}, {\it Astron. J. }{\bf 139}, 1438 (2010).

\bibitem{Wang2} S. Wang, X. -D. Li, and M. Li, {\it Phys. Rev. D }{\bf 82}, 103006 (2010).

\bibitem{Melia2} F. Melia, {\it Astron. Astrophys. }{\bf 553}, A76 (2013).

\end{thebibliography}
\end{document}